\documentclass[12pt]{article}
\usepackage[left=2.5cm,top=2.50cm,right=2.5cm,bottom=2.50cm]{geometry}
\usepackage{mathrsfs}
\usepackage{amsmath,amssymb,latexsym,color,cancel,graphicx,bbm,colortbl}
\usepackage[english]{babel}
\usepackage[latin1]{inputenc}
\usepackage{ragged2e}
\usepackage{cite}

\begin{document}
\date{}

\title{A Generalization of the Parametric Amplifier with Dunkl Derivative: Spectral and Statistical Properties}
\author{D. Ojeda-Guill\'en$^{a}$\footnote{{\it E-mail address:} dojedag@ipn.mx}, R. D. Mota$^{b}$, and J. C. Vega$^{c}$} \maketitle

\begin{minipage}{0.9\textwidth}

\small $^{a}$ Escuela Superior de C\'omputo, Instituto Polit\'ecnico Nacional,
Av. Juan de Dios B\'atiz esq. Av. Miguel Oth\'on de Mendiz\'abal, Col. Lindavista,
Alc. Gustavo A. Madero, C.P. 07738, Ciudad de M\'exico, Mexico.\\

\small $^{b}$ Escuela Superior de Ingenier{\'i}a Mec\'anica y El\'ectrica, Unidad Culhuac\'an,
Instituto Polit\'ecnico Nacional, Av. Santa Ana No. 1000, Col. San
Francisco Culhuac\'an, Alc. Coyoac\'an, C.P. 04430, Ciudad de M\'exico, Mexico.\\

\small $^{c}$ Escuela Superior de F{\'i}sica y Matem\'aticas, Instituto Polit\'ecnico Nacional,
Ed. 9, U.P. Adolfo L\'opez Mateos, Alc. Gustavo A. Madero, C.P. 07738 Ciudad de M\'exico, Mexico.\\

\end{minipage}

\begin{abstract}
We study the parametric amplifier Hamiltonian within the framework of the Dunkl formalism. We introduce the Dunkl creation and annihilation operators and show that their quadratic combinations generate an $su(1,1)$ Lie algebra. The spectral problem is solved exactly using two algebraic methods: the $su(1,1)$ tilting transformation and the generalized Bogoliubov transformation. The exact energy spectrum and the corresponding eigenfunctions are obtained in terms of the Dunkl number coherent states. Furthermore, we compute the Mandel $Q$ parameter and the second-order correlation function $g^{(2)}(0)$ to analyze the statistical properties of the Dunkl squeezed states. We show that, for the squeezed vacuum, the Mandel parameter remains independent of the Dunkl deformation, whereas the correlation function exhibits an explicit dependence on the Dunkl parameter $\mu$, which modifies the photon bunching effects. Finally, we show that our results reduce to the standard parametric amplifier case in the limit of vanishing Dunkl deformation parameter.
\end{abstract}

\section{Introduction}

The study of quantum mechanical systems through algebraic deformations has its roots in the seminal works of Wigner \cite{wigner1950}, who questioned whether the equations of motion determine the commutation relations uniquely. Shortly after, Yang \cite{yang1951} explored the quantization of space using reflection operators, laying the groundwork for deformations of the Heisenberg-Weyl algebra. Decades later, Dunkl introduced a differential-difference operator associated with reflection groups, now known as the Dunkl derivative \cite{dunkl1989,dunkl2001}. In recent years, the Dunkl derivative has emerged as a highly active area of research in mathematical physics. The harmonic oscillator and the Coulomb problem in two and three dimensions were the first physical systems to be successfully addressed within the Dunkl formalism \cite{genest2013,genest2014a,genest2014b,genest2015,nos2017,nos2018,ghazouani2019a,ghazouani2019b}. In these studies, the superintegrability of the models was established, and their exact solutions were constructed in terms of Dunkl spherical harmonics, revealing that the energy spectra exhibit a parity-dependent structure governed by $su(1,1)$ symmetry algebras.

Furthermore, the applicability of the Dunkl derivative has been successfully extended to the relativistic regime. In this context, the Dunkl-Klein-Gordon and Dunkl-Dirac oscillators have been exactly solved \cite{nos2019,nos2020, nos2021,hassanabadi2022,hamil2022}. These works demonstrate that the parity-dependent structure of the Dunkl derivative introduces non-trivial modifications to the energy spectra and wavefunctions in the relativistic domain as well.

More recently, numerous studies have successfully applied this formalism to generalize a wide array of quantum mechanical systems, extending its scope to exactly solvable oscillators, spatially dependent potentials, higher-dimensional relativistic wave equations, time-dependent dynamics, thermodynamic properties, and the construction of coherent states using algebraic methods \cite{quesne2023,junker2023,junker2024,schulze2024a,schulze2024b,schulze2024c,rabsaghari2024,hassanabadi2024,benzair2024,raber2025,benarous2025,hamil2025a,bouguerne2024a,bouguerne2024b,benchikha2024,lut2025,hocine2025,rou2023, hamil2023, hamil2025b, salazar2026}. These contributions demonstrate the effectiveness of the Dunkl formalism to reveal novel physical properties and parity structures, thereby motivating its application to more complex algebraic Hamiltonians.

On the other hand, the algebraic approach to quantum optics Hamiltonians has been a fundamental tool in the description of non-classical properties of light. Models such as the parametric amplifier and coupled oscillators have been extensively studied using $su(1,1)$ and $su(2)$ Lie algebras \cite{estes1968,caves1981,gerry1987}. These works established that algebraic methods allow for the diagonalization of Hamiltonians via tilting transformations and the exact calculation of the energy spectrum and Berry phases, as can be seen in Refs. \cite{chaturvedi1987,gerry1989}. Furthermore, the use of invariants and group-theoretical strategies has proven essential for understanding the dynamics of squeezed states and the geometric phase in generalized quantum systems \cite{vourdas1990}.

Currently, there is a renewed interest in extending these algebraic techniques to complex and deformed quantum systems. For instance, in Ref. \cite{villegas2018}, the authors developed algebraic methods to solve the master equation of the degenerate and nondegenerate parametric oscillator in the presence of a squeezed reservoir, while in Ref. \cite{makarov2018}, similar exact techniques were applied to time-dependent coupled quantum oscillators. More specific applications include the study of parity-deformed Tavis-Cummings models and their statistical properties \cite{algarni2022}, and the analysis of the time-dependent Mandel $Q$ parameter in photon sources \cite{jones2023}. In this context, the $su(1,1)$ algebraic framework has been successfully applied to solve the non-degenerate parametric amplifier \cite{nos2014}, and the $k$-photon Jaynes-Cummings model \cite{choreno2018}. More recently, this algebraic approach has been extended to compute the Berry phase and the Mandel parameter of various quantum optical Hamiltonians \cite{vega2024,vega2025}, demonstrating the applicability of these techniques in modern quantum optics.

Although the application of Dunkl-type operators in quantum optical contexts is not entirely novel, previous work has focused mainly on interaction models or static states. For instance, in Ref. \cite{moroz2016}, a connection is established between generalized Rabi models and Dunkl differential operators, while the relationship between relativistic Dunkl oscillators and the Anti-Jaynes-Cummings model is discussed in Ref. \cite{hamil2022b}. Similarly, the non-classical properties of coherent states associated with a parity-deformed boson algebra are investigated in Ref. \cite{chung2021}. However, the explicit study of the Dunkl parametric amplifier, which generates squeezed states, remains unexplored. The aim of this work is to construct the Dunkl generalization of this quadratic Hamiltonian. By replacing the standard creation and annihilation operators with their Dunkl-deformed counterparts, we obtain a solvable model that preserves the $su(1,1)$ algebraic structure while incorporating parity-dependent dynamics. This generalization allows us to investigate how the reflection symmetry inherent in the Dunkl derivative modifies the photon statistics, specifically the bunching and squeezing properties of the field, extending the results beyond previously studied coherent states.

This paper is organized as follows. In Section 2, we introduce the Dunkl $su(1,1)$ Lie algebra derived from the deformed creation and annihilation operators. In Section 3, we construct the Dunkl parametric amplifier Hamiltonian and demonstrate its commutation with the reflection operator, establishing the preservation of parity symmetry. Section 4 is dedicated to obtaining the exact solution of the Dunkl Hamiltonian using the tilting and generalized Bogoliubov transformations, followed by a spectral analysis via parity decomposition. In Section 5, we investigate the statistical properties of the system by calculating the Mandel $Q$ parameter and the second-order correlation function $g^{(2)}(0)$ for the general excited states. Finally, we present our concluding remarks in Section 6.

\section{The Dunkl $su(1,1)$ Lie Algebra}

The Dunkl derivative is a differential-difference operator associated with the reflection group $\mathbb{Z}_2$, which generalizes the standard derivative by including a parity-dependent singular term. This operator induces a deformation of the Heisenberg-Weyl algebra, providing a robust framework for constructing exactly solvable models that preserve the underlying Lie algebraic structures.

The Dunkl derivative is defined as \cite{dunkl1989,dunkl2001}
\begin{equation}
D_{\mu} = \frac{d}{dx} + \frac{\mu}{x}(1 - R),
\end{equation}
where $\mu$ is the Dunkl parameter and $R$ is the reflection operator acting on functions as $Rf(x) = f(-x)$. With this definition we can introduce the Dunkl creation and annihilation operators \cite{genest2013}
\begin{equation}
a_{\mu} = \sqrt{\frac{m\omega}{2\hbar}} x + \sqrt{\frac{\hbar}{2m\omega}} D_{\mu}, \quad\quad
a_{\mu}^\dagger = \sqrt{\frac{m\omega}{2\hbar}} x - \sqrt{\frac{\hbar}{2m\omega}} D_{\mu}.
\end{equation}
The Dunkl operators satisfy the following fundamental commutation relations
\begin{equation}
[a_{\mu}, a_{\mu}^\dagger] = 1 + 2\mu R,        \hspace{1.5cm} [a_{\mu}, R] = 2a_{\mu}R,    \hspace{1.5cm} [a_{\mu}^\dagger, R] = 2a_{\mu}^\dagger R.
\end{equation}
It is important to note that the last two relations imply the anti-commutation property $\{R, a_{\mu}\} = 0$.

The Dunkl number operator is defined as $N_{\mu} = a_{\mu}^\dagger a_{\mu}$ and satisfies the commutation relations
\begin{equation}
[N_{\mu}, a_{\mu}] = -a_{\mu}(1 - 2\mu R), \hspace{2cm} [N_{\mu}, a_{\mu}^\dagger] = a_{\mu}^\dagger(1 + 2\mu R).
\end{equation}

Now, we shall define the Dunkl quadratic operators as
\begin{equation}\label{Kdef}
K_+^{\mu} = \frac{1}{2}(a_{\mu}^\dagger)^2, \quad\quad
K_-^{\mu} = \frac{1}{2}a_{\mu}^2, \quad\quad
K_0^{\mu} = \frac{1}{4}(a_{\mu}^\dagger a_{\mu} + a_{\mu} a_{\mu}^\dagger).
\end{equation}

It can be proved that the operators $K_+^{\mu}, K_-^{\mu}, K_0^{\mu}$ satisfy the standard $su(1,1)$ Lie algebra relations. We observe that although the individual commutators involve the reflection operator, the anti-commutation property $\{R, a_{\mu}\} = 0$ ensures exact cancellation of all $\mu$-dependent terms in the bilinear products. This recovers the clean $su(1,1)$ structure.
For the commutator $[K_0^{\mu}, K_+^{\mu}]$ we have
\begin{equation}
[K_0^{\mu}, K_+^{\mu}] = \left[\frac{1}{4}(a_{\mu}^\dagger a_{\mu} + a_{\mu} a_{\mu}^\dagger), \frac{1}{2}(a_{\mu}^\dagger)^2\right]
= \frac{1}{2}(a_{\mu}^\dagger)^2 = K_+^{\mu}.
\end{equation}
Similarly, we can calculate the commutation relations between the operators $K_0^{\mu}$ and $K_-^{\mu}$ as follows
\begin{equation}
[K_0^{\mu}, K_-^{\mu}] = \left[\frac{1}{4}(a_{\mu}^\dagger a_{\mu} + a_{\mu} a_{\mu}^\dagger), \frac{1}{2}a_{\mu}^2\right]
= -\frac{1}{2}a_{\mu}^2 = -K_-^{\mu}.
\end{equation}
For the commutator $[K_-^{\mu}, K_+^{\mu}]$ the computation yields
\begin{equation}
[K_-^{\mu}, K_+^{\mu}] = \left[\frac{1}{2}a_{\mu}^2, \frac{1}{2}(a_{\mu}^\dagger)^2\right]
= \frac{1}{2}(a_{\mu}a_{\mu}^\dagger + a_{\mu}^\dagger a_{\mu}) = 2K_0^{\mu}.
\end{equation}
Thus, we have demonstrated the complete set of $su(1,1)$ commutation relations
\begin{equation}\label{Comm}
[K_0^{\mu}, K_+^{\mu}] = +K_+^{\mu}, \quad\quad [K_0^{\mu}, K_-^{\mu}] = -K_-^{\mu}, \quad\quad [K_-^{\mu}, K_+^{\mu}] = 2K_0^{\mu}.
\end{equation}

The action of the $su(1,1)$ generators on the Dunkl number states $|n\rangle$ can be explicitly written as
\begin{align}
    K_{0}^{\mu}|n\rangle &= \frac{1}{2}\left([n]_{\mu}+\frac{1}{2}+\mu(-1)^{n}\right)|n\rangle, \label{K0n} \\
    K_{+}^{\mu}|n\rangle &= \frac{1}{2}\sqrt{[n+1]_{\mu}[n+2]_{\mu}}|n+2\rangle, \label{K+n} \\
    K_{-}^{\mu}|n\rangle &= \frac{1}{2}\sqrt{[n]_{\mu}[n-1]_{\mu}}|n-2\rangle, \label{K-n}
\end{align}
where the Dunkl number is defined as $[n]_\mu = n + \mu(1-(-1)^n)$.
Furthermore, the Dunkl-Casimir operator, defined as $C^{\mu}=(K_{0}^{\mu})^{2}-K_{0}^{\mu}-K_{+}^{\mu}K_{-}^{\mu}$, commutes with all the generators of this $su(1,1)$ algebra. Its action on the Dunkl number states $|n\rangle$ is given by
\begin{equation}
    C^{\mu}|n\rangle = \left(\frac{\mu^{2}}{4}-\frac{\mu}{4}R-\frac{3}{16}\right)|n\rangle = k(k-1)|n\rangle.
\end{equation}
Hence, the Bargmann index $k_{+}$ (for the even sector $R=1$) and $k_{-}$ (for the odd sector $R=-1$) are written as
\begin{equation}
    k_{+}=\frac{1}{4}+\frac{\mu}{2}, \hspace{2cm} k_{-}=\frac{3}{4}+\frac{\mu}{2}.
\end{equation}

Despite the complicated commutation relation between the Dunkl creation and annihilation operators, the preservation of the $su(1,1)$ Lie algebra allows for the exact treatment of the model and demonstrates the suitability of this framework for quantum optical applications.

\section{The Dunkl Parametric Amplifier Hamiltonian}

In quantum optics, the parametric amplifier Hamiltonian describes a fundamental system where photon pairs are created or annihilated through nonlinear optical processes. The standard Hamiltonian is given by \cite{caves1981,gerry1989}
\begin{equation}
H = \omega a^\dagger a + f a^2 + f^* (a^\dagger)^2,
\end{equation}
where $\omega$ is the field frequency and $f$ is the complex pumping parameter characterizing the strength of the parametric process.
The Dunkl generalization of this Hamiltonian replaces the standard creation and annihilation operators with their Dunkl-deformed counterparts
\begin{equation}\label{Hu}
H_{\mu} = \omega a_{\mu}^\dagger a_{\mu} + f a_{\mu}^2 + f^* (a_{\mu}^\dagger)^2.
\end{equation}
This Hamiltonian models parametric amplification in systems with additional reflection symmetry properties encoded through the Dunkl formalism.

Using the Dunkl $su(1,1)$ generators defined previously, we can express the Hamiltonian in a particularly convenient form. First, we recall the relations
\begin{equation}
N_\mu=a_{\mu}^\dagger a_{\mu} = 2K_0^{\mu} - \frac{1}{2}-\mu R, \hspace{1.5cm} a_{\mu}^2 = 2K_{-}^{\mu}, \hspace{1.5cm} (a_{\mu}^\dagger)^2 = 2K_{+}^{\mu}.
\end{equation}
Substituting these results into Eq. (\ref{Hu}) we obtain (up to an additive constant)
\begin{equation}\label{Ham}
H_{\mu}=2\omega K_0^{\mu} + 2f K_{-}^{\mu} + 2f^* K_{+}^{\mu}.
\end{equation}
The Dunkl number operator $N_{\mu} = a_{\mu}^\dagger a_{\mu}$ commutes with the Hamiltonian as follows
\begin{equation}
[H_{\mu}, N_{\mu}] = 4f K_{-}^\mu - 4f^* K_{+}^\mu = 2f a_{\mu}^2 - 2f^* (a_{\mu}^\dagger)^2.
\end{equation}
Thus, the number of excitations is not conserved in the parametric amplification process, consistent with the physical picture of photon pair generation.

A crucial property of the Dunkl Hamiltonian $H_\mu$ is its behavior under reflection. The commutation with the reflection operator $R$ is given by
\begin{equation}
[H_{\mu}, R] = 2\omega[K_0^\mu, R] + 2f[K_{-}^\mu, R] + 2f^*[K_{+}^\mu, R] = 0.
\end{equation}
This demonstrates that the reflection symmetry is preserved by the Dunkl parametric amplifier Hamiltonian, making $R$ a constant of motion. Consequently, the Hilbert space splits into two independent sectors that do not mix under time evolution: the even sector (associated with even parity states) and the odd sector (associated with odd parity states). This separation justifies the existence of two distinct Bargmann indices $k_\pm$ depending on the eigenvalue of $R=\pm 1$, allowing the system to be solved independently for each parity sector. These results complete our analysis of the algebraic properties of the Hamiltonian, setting the stage for the dynamical study of the system.

\section{Exact Solution and Diagonalization of the Dunkl parametric amplifier Hamiltonian}

In this section, we obtain the exact solution of the Dunkl parametric amplifier problem using two complementary algebraic approaches: the $su(1,1)$ tilting transformation and the generalized Bogoliubov transformation. Furthermore, we explicitly analyze the parity structure of the system to determine the splitting of the energy spectrum in the even and odd sectors.

\subsection{The $su(1,1)$ tilting transformation}

To obtain the energy spectrum and eigenfunctions of the Hamiltonian given in Eq. (\ref{Ham}),
\begin{equation}\label{Ham2}
H_{\mu} = 2\omega K_0^{\mu} + 2f K_{-}^{\mu} + 2f^* K_{+}^{\mu}.
\end{equation}
we apply the tilting transformation to the stationary Schr\"odinger equation $H_{\mu}\Psi=E\Psi$. This method allows us to diagonalize the Hamiltonian by applying a unitary rotation operator. We introduce the tilted Hamiltonian $H'_{\mu}$ and the transformed wavefunction $\Psi'$, in the same way that it has been used to study Hamiltonians of quantum optics, defined with the standard derivative as follows \cite{gerry1989,nos2014,nos2019b,nos2021b}
\begin{equation}\label{Ht}
H'_{\mu} = D^{\dagger}(\xi) H_{\mu} D(\xi), \hspace{1.5cm} \Psi' = D^{\dagger}(\xi)\Psi,
\end{equation}
where $D(\xi)$ is the $su(1,1)$ displacement operator defined as
\begin{equation}
    D(\xi)=\exp(\xi K_{+}^{\mu}-\xi^{*}K_{-}^{\mu}),
\end{equation}
with $\xi=-\frac{1}{2}\tau e^{-i\varphi}$, $-\infty<\tau<\infty$ and $0\leq\varphi\leq2\pi$. By using the Baker-Campbell-Hausdorff identity and the commutation relations of the Dunkl $su(1,1)$ algebra derived in Section 2, we can compute the similarity transformations for the $su(1,1)$ generators to obtain
\begin{eqnarray}
\tilde{K}_{0}^{\mu} &=& D^{\dagger}(\xi)K_{0}^{\mu}D(\xi)=\cosh(\tau)K_{0}^{\mu}-\frac{1}{2}\sinh(\tau)e^{-i\varphi}K_{+}^{\mu}-\frac{1}{2}\sinh(\tau)e^{i\varphi}K_{-}^{\mu}, \label{K_0t}\\
\tilde{K}_{+}^{\mu} &=& D^{\dagger}(\xi)K_{+}^{\mu}D(\xi)=-\sinh(\tau)e^{i\varphi}K_{0}^{\mu}+\cosh^{2}\left(\frac{\tau}{2}\right)K_{+}^{\mu}+\sinh^{2}\left(\frac{\tau}{2}\right)e^{2i\varphi}K_{-}^{\mu}, \label{K_+t}\\
\tilde{K}_{-}^{\mu} &=& D^{\dagger}(\xi)K_{-}^{\mu}D(\xi)=-\sinh(\tau)e^{-i\varphi}K_{0}^{\mu}+\sinh^{2}\left(\frac{\tau}{2}\right)e^{-2i\varphi}K_{+}^{\mu}+\cosh^{2}\left(\frac{\tau}{2}\right)K_{-}^{\mu}. \label{K_-t}
\end{eqnarray}

Substituting these results into the expression for $H_{\mu}$ of Eq. (\ref{Ht}) and grouping the terms associated with the operators $K_{0}^{\mu}$, $K_{+}^{\mu}$, and $K_{-}^{\mu}$, we obtain the explicit form of the tilted Hamiltonian
\begin{align}\label{Ht2}
    H_{\mu}^{\prime} &= [2\omega\cosh(\tau)-2f\sinh(\tau)e^{-i\varphi}-2f^{*}\sinh(\tau)e^{i\varphi}]K_{0}^{\mu} \nonumber \\
    &\quad + \left[-\omega\sinh(\tau)e^{-i\varphi}+2f\sinh^{2}\left(\frac{\tau}{2}\right)e^{-2i\varphi}+2f^{*}\cosh^{2}\left(\frac{\tau}{2}\right)\right]K_{+}^{\mu} \nonumber \\
    &\quad + \left[-\omega\sinh(\tau)e^{i\varphi}+2f\cosh^{2}\left(\frac{\tau}{2}\right)+2f^{*}\sinh^{2}\left(\frac{\tau}{2}\right)e^{2i\varphi}\right]K_{-}^{\mu}.
\end{align}
To diagonalize this Hamiltonian, we require the coefficients of the non-diagonal operators $K_{+}^{\mu}$ and $K_{-}^{\mu}$ to vanish. To explicitly incorporate the phase dependency, we write the complex pumping parameter as $f = |f|e^{i\theta}$, where $\theta$ is the phase of the pump. Following the algebraic approach of Refs. \cite{gerry1989,nos2014,nos2019b,nos2021b}, we eliminate the non-diagonal terms by setting the phase of the coherent state equal to the phase of the pump,
\begin{equation}
    \varphi = \theta.
\end{equation}
Under this choice, all terms in the coefficient of $K_{+}^{\mu}$ become proportional to $e^{-i\theta}$, and the condition for them to vanish reduces to the real algebraic equation
\begin{equation}
    -\omega\sinh(\tau)+2|f|\left[\sinh^{2}\left(\frac{\tau}{2}\right)+\cosh^{2}\left(\frac{\tau}{2}\right)\right] = 0.
\end{equation}
Using the hyperbolic identity $\cosh^2(\tau/2) + \sinh^2(\tau/2) = \cosh(\tau)$, this simplifies directly to the condition for the squeezing parameter
\begin{equation}
    \tanh(\tau)=\frac{2|f|}{\omega}.
\end{equation}

It is important to note that this choice is valid provided that the stability condition $\omega > 2|f|$ holds (see equation (\ref{HtK})). With these parameters, the non-diagonal terms are eliminated, and the tilted Hamiltonian reduces to a diagonal form proportional to the generator $K_0^{\mu}$
\begin{equation}\label{HtK}
H'_{\mu} = 2\sqrt{\omega^2 - 4|f|^2} K_0^{\mu} = \Omega_{\mu} K_0^{\mu},
\end{equation}
where we have defined the generalized Rabi frequency $\Omega_{\mu} = 2\sqrt{\omega^2 - 4|f|^2}$. Since the operator $K_0^{\mu}$ is diagonal in the Fock basis, we can immediately obtain the energy spectrum by evaluating its action on the Dunkl number states, as given in Eq. (\ref{K0n})
\begin{equation}
E_n = \frac{1}{2} \Omega_{\mu} \left( [n]_{\mu} + \frac{1}{2} + \mu(-1)^n \right).
\end{equation}
By substituting the explicit form of the Dunkl number $[n]_{\mu} = n + \mu(1 - (-1)^n)$, the alternating terms exactly cancel out, leading to the unified energy spectrum
\begin{equation}\label{Espec1}
E_n = \frac{1}{2} \Omega_{\mu} \left( n + \mu + \frac{1}{2} \right), \quad n=0, 1, 2, \dots
\end{equation}
The eigenfunctions of the original Hamiltonian $H_{\mu}$ are obtained by applying the displacement operator to the standard number states, yielding the Dunkl $su(1,1)$ Perelomov number coherent states
\begin{equation}
|\zeta, n\rangle = D(\xi) |n\rangle = |\zeta, n\rangle,
\end{equation}
where $\zeta = -\tanh(\frac{\tau}{2})e^{-i\varphi}$.
The explicit algebraic expansion of these Dunkl Perelomov number coherent states in the position representation, $\Psi_n(x) = \langle x | \zeta, n \rangle$, can be obtained directly using the general summation formulas derived in Ref. \cite{nos2014}.

It is important to highlight that in the standard limit $\mu \rightarrow 0$, the Dunkl derivative reduces to the standard derivative, and the deformed algebra recovers the standard Heisenberg-Weyl structure. Consequently, our results for the energy spectrum and the diagonalized Hamiltonian exactly reproduce the well-known results for the degenerate parametric amplifier obtained via the tilting transformation by Gerry \cite{gerry1989}. Moreover, the spatial wavefunctions reduce exactly to the well-known squeezed number states of the one-dimensional harmonic oscillator. As it was demonstrated by Nieto \cite{nieto1997}, these standard squeezed states are expressed in terms of the standard Hermite polynomials $H_n(x)$ and specific functions of the coherent state parameters. This consistency confirms that the Dunkl formalism provides a natural generalization of the standard quantum optical models, preserving the fundamental algebraic properties while introducing parity-dependent features.

\subsection{The Generalized Bogoliubov Transformation}

Alternatively, we can diagonalize the Hamiltonian of Eq. (\ref{Ham2}),
\begin{equation}
H_{\mu} = 2\omega K_0^{\mu} + 2f K_{-}^{\mu} + 2f^* K_{+}^{\mu},
\end{equation}
by introducing a generalized Bogoliubov transformation that preserves the deformed commutation relations. We seek a linear transformation of the Dunkl creation and annihilation operators of the form \cite{caves1981}
\begin{equation}
b_{\mu} = u a_{\mu} + v a_{\mu}^{\dagger}, \hspace{2cm} b_{\mu}^{\dagger} = u^* a_{\mu}^{\dagger} + v^* a_{\mu},
\end{equation}
where $u$ and $v$ are complex coefficients. Solving the system of equations for the original operators $a$ and $a^{\dagger}$, we obtain the inverse transformation
\begin{equation}\label{Invbog}
a_{\mu} = u^* b_{\mu} - v b_{\mu}^{\dagger}, \hspace{2cm} a_{\mu}^{\dagger} = u b_{\mu}^{\dagger} - v^* b_{\mu},
\end{equation}
For the new operators to satisfy the same fundamental Dunkl commutation relation $[b_{\mu}, b_{\mu}^{\dagger}] = 1 + 2\mu R$, which ensures that the Bogoliubov transformation is canonical and preserves the underlying $su(1,1)$ algebraic structure of the Dunkl formalism, the coefficients must satisfy the condition $|u|^2 - |v|^2 = 1$. This guarantees that the new operators $b_{\mu}$ and $b_{\mu}^{\dagger}$ act as valid quasiparticle creation and annihilation operators within the same Hilbert space. This allows us to parameterize them as $u=\cosh(r)$ and $v=e^{-i\phi}\sinh(r)$, where $\phi$ is the phase of the Bogoliubov transformation. Substituting the inverse transformation into the original Hamiltonian $H_{\mu}$ and regrouping terms, we obtain
\begin{equation}
    H_{\mu}^{\prime\prime}=\mathcal{A}b_{\mu}^{\dagger}b_{\mu}+\mathcal{B}b_{\mu}^{2}+\mathcal{B}^{*}(b_{\mu}^{\dagger})^{2}+\mathcal{C},
\end{equation}
where the coefficients are defined as
\begin{align}
    \mathcal{A} &= \omega(u^{2}+|v|^{2})-2fuv-2f^{*}uv^{*}, \\
    \mathcal{B} &= fu^{2}+f^{*}(v^{*})^{2}-\omega uv^{*}, \\
    \mathcal{C} &= \omega|v|^{2}-fuv-f^{*}uv^{*}.
\end{align}

Now, we require the coefficients of the terms proportional to $b_{\mu}^{2}$ and $(b_{\mu}^{\dagger})^{2}$ to vanish. This leads to the condition $\mathcal{B}=0$, which explicitly gives
\begin{equation}
    fu^{2}+f^{*}(v^{*})^{2}-\omega uv^{*}=0.
\end{equation}
By using the pumping parameter $f=|f|e^{i\theta}$ and our parameterization for $u$ and $v$, this condition becomes
\begin{equation}
    |f|e^{i\theta}\cosh^{2}(r)+|f|e^{-i\theta}(e^{i\phi}\sinh(r))^{2}-\omega\cosh(r)(e^{i\phi}\sinh(r))=0.
\end{equation}
To eliminate the phase dependence and satisfy the equation with real parameters, we set the Bogoliubov phase equal to the pump phase,
\begin{equation}
    \phi = \theta.
\end{equation}
Under this choice, all terms become proportional to $e^{i\theta}$, allowing the phases to cancel out perfectly. This yields the real algebraic equation
\begin{equation}
    |f|(\cosh^{2}(r)+\sinh^{2}(r))-\omega\cosh(r)\sinh(r)=0,
\end{equation}
which simplifies to the standard condition for the squeezing parameter
\begin{equation}
    \tanh(2r)=\frac{2|f|}{\omega}.
\end{equation}
It is worth noting that the Bogoliubov squeezing parameter $r$ is directly related to the tilting parameter $\tau$ used in the previous subsection by the relation $\tau = 2r$. Consequently, the diagonalization condition $\tanh(2r) = 2|f|/\omega$ is mathematically equivalent to $\tanh(\tau) = 2|f|/\omega$, ensuring the consistency between both algebraic approaches.

Under this transformation, the Hamiltonian takes the diagonal form of a free Dunkl oscillator with a renormalized frequency
\begin{equation}
H''_{\mu} = \frac{1}{2}\Omega_{\mu} \left( b_{\mu}^{\dagger} b_{\mu} + \frac{1}{2} + \mu R \right),
\end{equation}
where we use the generalized Rabi frequency $\Omega_{\mu} = 2\sqrt{\omega^2 - 4|f|^2}$ previously defined. The term $b_{\mu}^{\dagger} b_{\mu}$ is the number operator in the transformed basis, which also has the Dunkl numbers $[n]_\mu = n + \mu(1 - (-1)^n)$ as eigenvalues, with $n=0,1,2,\dots$. The reflection operator $R$ commutes with the Hamiltonian and has eigenvalues $(-1)^n$. By substituting these eigenvalues into the diagonal Hamiltonian and expanding the Dunkl number, the alternating terms cancel out exactly, yielding the energy spectrum
\begin{equation}
E_n = \frac{1}{2}\Omega_{\mu} \left( [n]_\mu + \frac{1}{2} + \mu (-1)^n \right) = \frac{1}{2}\Omega_{\mu} \left( n + \mu + \frac{1}{2} \right).
\end{equation}
This result demonstrates that the generalized Bogoliubov transformation exactly matches the unified energy spectrum obtained via the $su(1,1)$ tilting transformation in Eq. (\ref{Espec1}), confirming the absolute consistency of both algebraic approaches.

\subsection{Spectrum Analysis and Parity Decomposition}

With the Hamiltonian diagonalized, we now analyze the specific structure of the energy spectrum imposed by the Dunkl deformation. A rigorous feature of the Dunkl formalism is the preservation of parity symmetry. Since $[H_{\mu}, R] = 0$, the Hilbert space $\mathcal{H}$ decomposes into two orthogonal subspaces: the even sector $\mathcal{H}_+$ (where $R=+1$) and the odd sector $\mathcal{H}_-$ (where $R=-1$). This allows us to split the unified energy spectrum obtained in Eq. (\ref{Espec1}) into two distinct branches depending on the parity of the number state $|n\rangle$.

In the even sector ($\mathcal{H}_+$), we consider states with even occupation number $n=2m$ ($m=0,1,\dots$). Substituting this directly into the general energy formula, we obtain the spectrum for the even states
\begin{equation}
E_{2m} = \frac{1}{2}\Omega_{\mu} \left( 2m + \mu + \frac{1}{2} \right).
\end{equation}

In the odd sector ($\mathcal{H}_-$), we consider states with odd occupation number $n=2m+1$. The corresponding energy spectrum for the odd states is obtained by substituting this into the general formula as follows
\begin{equation}
E_{2m+1} = \frac{1}{2}\Omega_{\mu} \left( 2m + \mu + \frac{3}{2} \right).
\end{equation}

This analysis reveals that the Dunkl parameter $\mu$ lifts the degeneracy of the energy spacing. While the dynamical frequency $\Omega_{\mu}$ is determined exclusively by the pumping parameters $\omega$ and $f$, the ``zero-point" energy and the separation between consecutive levels are modified by $\mu$ in a parity-dependent manner. This is the unique effect of the Dunkl deformation in the parametric amplifier.

Therefore, we have demonstrated that the Dunkl parametric amplifier can be exactly solved using algebraic methods. The tilting transformation provides the direct route to the wavefunctions, while the Bogoliubov approach offers a quasiparticle interpretation. Our analysis of the parity sectors shows that the Dunkl deformation introduces a parity-dependent shift in the spectrum without altering the fundamental stability condition $\omega > 2|f|$.

\section{Photon Statistics of the Dunkl Parametric Amplifier}

In this section, we analyze the statistical properties of the system, specifically the Mandel $Q$ parameter and the second-order correlation function $g^{(2)}(0)$. Instead of using the inverse Bogoliubov transformation for individual operators, we employ the $su(1,1)$ algebraic approach (similarity transformations) to calculate the expectation values directly.

The similarity transformations of the $su(1,1)$ generators under the displacement operator $D(\xi)$, denoted as $\tilde{K}_{i}^{\mu}$, were previously obtained in Eqs. (\ref{K_0t})-(\ref{K_-t}). We calculate the expectation values with respect to the Dunkl number states $|n\rangle$, which are eigenstates of $K_{0}^{\mu}$ as shown in Eq. (\ref{K0n}). Since the states $|n\rangle$ are orthogonal, the expectation values of the ladder operators vanish, i.e., $\langle n|K_{\pm}^{\mu}|n\rangle=0$. Thus, only the diagonal terms in the transformed generators contribute to the statistics.

\subsection{Mandel $Q$ Parameter}

The Mandel $Q$ parameter characterizes the photon number distribution and is defined as \cite{mandel1979,mandel1995}
\begin{equation}
Q_{n}^{\mu}=\frac{\langle(\Delta N_{\mu})^{2}\rangle}{\langle N_{\mu}\rangle}-1=\frac{\langle N_{\mu}^{2}\rangle-\langle N_{\mu}\rangle^{2}}{\langle N_{\mu}\rangle}-1.
\end{equation}
The Dunkl number operator $N_\mu$ can be written in terms of the $K_0^\mu$ generator as $N_\mu = 2K_0^\mu - \frac{1}{2} - \mu R$. Since the reflection operator $R$ commutes with the quadratic generators $K_\pm^\mu$, it remains invariant under the transformation $D(\xi)$.

Using Eq. (\ref{K_0t}) and since $\langle K_\pm^\mu \rangle = 0$, we only obtain contributions from the diagonal terms. Therefore,
\begin{align}
    \langle N_\mu \rangle &= \langle n | D^\dagger (2K_0^\mu - 1/2 - \mu R) D | n \rangle \nonumber \\
    &= 2\cosh(2r)\langle n | K_0^\mu | n \rangle - \frac{1}{2} - \mu \langle n | R | n \rangle.
\end{align}
Substituting the eigenvalue $\lambda_n$ and $\langle R \rangle = (-1)^n$
\begin{equation}
    \langle N_\mu \rangle = \cosh(2r)([n]_\mu + \frac{1}{2} + \mu(-1)^n) - \frac{1}{2} - \mu(-1)^n.
\end{equation}
Using the identity $\cosh(2r) = \cosh^2(r) + \sinh^2(r)$ and the properties of Dunkl numbers, this simplifies to
\begin{equation} \label{N}
    \langle N_\mu \rangle = \cosh^2(r)[n]_\mu + \sinh^2(r)[n+1]_\mu.
\end{equation}

In order to find $\langle N_{\mu}^{2}\rangle$, we compute the expectation value of $(2\tilde{K}_{0}^{\mu}-C_{R})^{2}$ where $C_{R}=1/2+\mu R$. The square of the transformed generator $(\tilde{K}_{0}^{\mu})^{2}$ contains terms like $(K_{0}^{\mu})^{2}$ and $K_{+}^{\mu}K_{-}^{\mu}$. Again, since only the diagonal terms contribute, it follows that
\begin{equation}
\langle(\tilde{K}_{0}^{\mu})^{2}\rangle=\cosh^{2}(2r)\langle(K_{0}^{\mu})^{2}\rangle+\frac{1}{4}\sinh^{2}(2r)\langle\{K_{+}^{\mu},K_{-}^{\mu}\}\rangle.
\end{equation}
Using the anticommutator relation $\{K_+^\mu, K_-^\mu\} = 2(K_0^\mu)^2 - 2C^\mu$, where $C^\mu$ is the Casimir invariant, we derive the variance $(\Delta N_\mu)^2 = \langle N_\mu^2 \rangle - \langle N_\mu \rangle^2$. After algebraic simplification using hyperbolic identities, we obtain
\begin{equation}
    (\Delta N_\mu)^2 = \frac{1}{4}\sinh^2(2r) \left( [n]_\mu [n-1]_\mu + [n+1]_\mu [n+2]_\mu \right).
\end{equation}
Finally, the Mandel parameter is given by
\begin{equation}\label{Mandelfin}
Q_{n}^{\mu}=\frac{\sinh^{2}(2r)}{4}\left(\frac{[n]_{\mu}[n-1]_{\mu}+[n+1]_{\mu}[n+2]_{\mu}}{\cosh^{2}(r)[n]_{\mu}+\sinh^{2}(r)[n+1]_{\mu}}\right)-1.
\end{equation}
This general expression reduces to the result $Q_{0}^{\mu}=\cosh(2r)$ for the ground state ($n=0$, where $[0]_{\mu}=0$ and $[1]_{\mu}=1+2\mu$), showing that the Mandel parameter for the squeezed vacuum is completely independent of the Dunkl deformation $\mu$ and consistently exhibits super-Poissonian statistics.

For excited states ($n \ge 1$), the Mandel parameter exhibits a strong dependence on the Dunkl deformation. It is instructive to verify that our result reduces to the known expression for standard squeezed number states when this deformation vanishes. In the limit $\mu\rightarrow 0$, the Dunkl number becomes the standard integer, $[n]_{\mu}\rightarrow n$. Substituting this into Eq. (66), we obtain the explicit expression for the standard parametric amplifier
\begin{equation}
Q_{n}^{(\mu=0)} = \frac{\sinh^2(2r)}{4}\left(\frac{n(n-1) + (n+1)(n+2)}{n\cosh^2(r) + (n+1)\sinh^2(r)}\right) - 1.
\end{equation}
This expression exactly recovers the well-known photon statistics for standard squeezed number states \cite{kim1989}, confirming the overall consistency of our generalized formalism across all energy levels.

\subsection{Second-Order Correlation Function}

The second-order correlation function $g^{(2)}(0)$ characterizes the intensity correlations and is defined as \cite{mandel1995,glauber2007}
\begin{equation}
g_{\mu}^{(2)}(0)=\frac{\langle(a_{\mu}^{\dagger})^{2}a_{\mu}^{2}\rangle}{\langle N_{\mu}\rangle^{2}}.
\end{equation}
To evaluate the numerator, we express the operators in terms of the $su(1,1)$ generators. Using the relations $a_{\mu}^{2}=2K_{-}^{\mu}$ and $(a_{\mu}^{\dagger})^{2}=2K_{+}^{\mu}$ the operator in the numerator is identified as $4K_{+}^{\mu}K_{-}^{\mu}$. Working in the Heisenberg picture, we calculate the expectation value of the transformed operator $\langle4\tilde{K}_{+}^{\mu}\tilde{K}_{-}^{\mu}\rangle$.

By applying the similarity transformations to the generators and multiplying $\tilde{K}_{+}^{\mu}$ and $\tilde{K}_{-}^{\mu}$, we observe that due to the orthogonality of the Fock states, only the terms that preserve the photon number contribute to the expectation value. Keeping only these diagonal contributions $((K_{0}^{\mu})^{2},K_{+}^{\mu}K_{-}^{\mu},K_{-}^{\mu}K_{+}^{\mu})$ the expectation value expands explicitly as
\begin{equation}
\langle4\tilde{K}_{+}^{\mu}\tilde{K}_{-}^{\mu}\rangle=4\sinh^{2}(2r)\langle(K_{0}^{\mu})^{2}\rangle+4\cosh^{4}(r)\langle K_{+}^{\mu}K_{-}^{\mu}\rangle+4\sinh^{4}(r)\langle K_{-}^{\mu}K_{+}^{\mu}\rangle.
\end{equation}
We evaluate these expectation values with respect to the Dunkl number states $|n\rangle$. For the quadratic generator term, using the eigenvalue $\lambda_n = \frac{1}{2}([n]_\mu + 1/2 + \mu(-1)^n)$ and the identity $[n]_\mu = n + \mu(1-(-1)^n)$, we obtain
\begin{equation}
    4\langle (K_0^\mu)^2 \rangle = \frac{1}{4}(2n+1+2\mu)^2.
\end{equation}
For the products of the ladder operators, acting on the states yields
\begin{align}
    4\langle K_+^\mu K_-^\mu \rangle &= [n]_\mu [n-1]_\mu, \\
    4\langle K_-^\mu K_+^\mu \rangle &= [n+1]_\mu [n+2]_\mu.
\end{align}
Substituting these explicit results into the numerator, and dividing by the squared mean photon number $\langle N_{\mu}\rangle^{2}=(\cosh^{2}(r)[n]_{\mu}+\sinh^{2}(r)[n+1]_{\mu})^{2}$ derived in the previous section, we arrive at the general expression for the correlation function
\begin{equation}\label{g2final}
g_{\mu}^{(2)}(0)=\frac{\cosh^{4}(r)[n]_{\mu}[n-1]_{\mu}+\frac{1}{4}\sinh^{2}(2r)(2n+1+2\mu)^{2}+\sinh^{4}(r)[n+1]_{\mu}[n+2]_{\mu}}{(\cosh^{2}(r)[n]_{\mu}+\sinh^{2}(r)[n+1]_{\mu})^{2}}.
\end{equation}
Hence, the second-order correlation function strongly depends on the Dunkl parameter $\mu$. This dependence appears not only through the generalized Dunkl numbers but also explicitly in the term $(2n+1+2\mu)^2$. Consequently, unlike the Mandel parameter for the squeezed vacuum, the intensity correlations are directly modified by the deformation. This demonstrates that the Dunkl parameter acts as a theoretical tuning mechanism that modifies the photon bunching effects of the field.

Finally, we examine the limiting cases to verify the consistency of our result. First, we consider the standard quantum mechanical limit by setting $\mu \to 0$. In this limit, the Dunkl number reduces to the standard integer $[n]_{\mu} \to n$, and the term $(2n + 1 + 2\mu)$ simplifies to $(2n+1)$. Consequently, Eq. (\ref{g2final}) reduces to the standard expression for squeezed number states
\begin{equation}
    g_{n}^{(2)}(0)\big|_{\mu=0} = \frac{n(n-1)\cosh^4(r) + \frac{1}{4}(2n+1)^2\sinh^2(2r) + (n+1)(n+2)\sinh^4(r)}{\left( n\cosh^2(r) + (n+1)\sinh^2(r) \right)^2}.
\end{equation}
From this result, we can straightforwardly obtain the correlation function for the standard squeezed vacuum by setting $n=0$. In this case, the first term in the numerator vanishes, and the expression simplifies to
\begin{equation}
    g_{0}^{(2)}(0)\big|_{\mu=0} = \frac{\frac{1}{4}\sinh^2(2r) + 2\sinh^4(r)}{\sinh^4(r)} = \frac{\cosh^2(r)\sinh^2(r) + 2\sinh^4(r)}{\sinh^4(r)}.
\end{equation}
Using the identity $\coth^2(r) + 2 = 3 + \frac{1}{\sinh^2(r)}$, we exactly recover the well-known result
\begin{equation}
    g_{0}^{(2)}(0)\big|_{\mu=0} = \coth^2(r) + 2 \ge 3.
\end{equation}
This confirms that our generalized formula correctly reproduces the super-Poissonian statistics and photon bunching characteristic of the standard parametric amplifier in the absence of Dunkl deformation \cite{walls1994}.

To conclude this section, it is important to remark that the statistical results derived using the $su(1,1)$ similarity transformations are entirely consistent with those obtained from the generalized Bogoliubov transformation. If one uses the inverse Bogoliubov relations defined in Section 4.2 to express the original operators in terms of the quasiparticle operators, the direct evaluation of the expectation values produces exactly the same analytical expressions for both the Mandel parameter $Q_n$ and the second-order correlation function $g^{(2)}(0)$. This equivalence confirms the validity of our results and demonstrates that both algebraic approaches provide a unified and consistent description of the statistical properties of the Dunkl parametric amplifier.

\section{Concluding remarks}

In this paper, we have investigated the Dunkl generalization of the parametric amplifier Hamiltonian. By introducing the Dunkl creation and annihilation operators, we demonstrated that their quadratic combinations strictly preserve the $su(1,1)$ Lie algebra. This algebraic structure allowed us to solve the spectral problem exactly through two independent approaches: the $su(1,1)$ tilting transformation and the generalized Bogoliubov transformation. Both methods consistently produced the exact energy spectrum and the corresponding eigenfunctions in terms of the Dunkl number coherent states. We showed that the Dunkl parameter $\mu$ introduces a parity-dependent shift in the energy levels, lifting the degeneracy while maintaining the fundamental stability condition of the system.

Furthermore, we analyzed the statistical properties of the Dunkl squeezed states by computing the Mandel $Q$ parameter and the second-order correlation function $g^{(2)}(0)$. Our results demonstrate that, specifically for the squeezed vacuum, the Mandel parameter remains completely independent of the deformation, consistently exhibiting super-Poissonian statistics. In contrast, the correlation function strongly depends on the Dunkl parameter across all states, providing a theoretical mechanism to tune the photon bunching effects. Finally, we verified that in the limit $\mu\rightarrow 0$, all the derived analytical expressions correctly reduce to the well-known standard quantum optical results, including the spatial wavefunctions, which exactly recover the standard squeezed number states expressed in terms of Hermite polynomials.

\section*{Acknowledgments}

This work was partially supported by SNII-M\'exico, COFAA-IPN, EDI-IPN, and CGPI-IPN Project Number 20251355.\\

\section*{Disclosures}

The authors declare no conflicts of interest.

\section*{Data Availability}

No data were generated or analyzed in the presented research.


\begin{thebibliography} {99}

\bibitem{wigner1950}E.P. Wigner, {\it Phys. Rev.} \textbf{77} (1950) 711.

\bibitem{yang1951}L.M. Yang, {\it Phys. Rev.} \textbf{84} (1951) 788.

\bibitem{dunkl1989} C.F. Dunkl, {\it Trans. Am. Math. Soc.} \textbf{311} (1989) 167.

\bibitem{dunkl2001}C.F. Dunkl, and Y. Xu, {\it Orthogonal polynomials of several variables}, Cambridge University Press, Cambridge, 2014.

\bibitem{genest2013}V.X. Genest, M.E.H. Ismail, L. Vinet, and A. Zhedanov, {\it J. Phys. A.} \textbf{46} (2013) 145201.

\bibitem{genest2014a}V.X. Genest, M.E.H. Ismail, L. Vinet, and A. Zhedanov, {\it Commun. Math. Phys.} {\bf329} (2014) 999.

\bibitem{genest2014b}V.X. Genest, L. Vinet, and A. Zhedanov, {\it J. Phys. Conf. Ser.} \textbf{512} (2014) 012010.

\bibitem{genest2015}V.X. Genest, A. Lapointe, and L. Vinet, {\it Phys. Lett. A} \textbf{379} (2015) 923.

\bibitem{nos2017}M. Salazar-Ram\'{\i}rez, D. Ojeda-Guill\'en, R.D. Mota, and V.D. Granados, {\it Eur. Phys. J. Plus} \textbf{132} (2017) 39.

\bibitem{nos2018}M. Salazar-Ram\'{\i}rez, D. Ojeda-Guill\'en, R.D. Mota, and V.D. Granados, {\it Mod. Phys. Lett. A} \textbf{33} (2018) 1850112.

\bibitem{ghazouani2019a}S. Ghazouani, I. Sboui, M.A. Amdouni, and M.B. El Hadj Rhouma, {\it J. Phys. A: Math. Theor.} \textbf{52} (2019) 225202.

\bibitem{ghazouani2019b}S. Ghazouani, and I. Sboui, {\it J. Phys. A: Math. Theor.} \textbf{53} (2019) 035202.



\bibitem{nos2019}R.D. Mota, D. Ojeda-Guill\'en, M. Salazar-Ram\'irez, and V.D. Granados, {\it Ann. Phys.} \textbf{411} (2019) 167964.

\bibitem{nos2020}D. Ojeda-Guill\'en, R.D. Mota, M. Salazar-Ram\'irez, and V.D. Granados, {\it Mod. Phys. Lett. A} \textbf{35} (2020) 2050255.

\bibitem{nos2021}R.D. Mota, D. Ojeda-Guill\'en, M. Salazar-Ram\'irez, and V.D. Granados, {\it  Mod. Phys. Lett. A} \textbf{36} (2021) 2150171.

\bibitem{hassanabadi2022}S. Hassanabadi, J. K\v{r}\'i\v{z}, B.C. L\"utf\"uo\u{g}lu and H. Hassanabadi, {\it Phys. Scr.} \textbf{97} (2022) 125305.

\bibitem{hamil2022}B. Hamil and B.C. L\"utf\"uo\u{g}lu, {\it Few-Body Syst.} \textbf{63} (2022) 74.




\bibitem{quesne2023} C. Quesne, {\it J. Phys. A: Math. Theor.} \textbf{56} (2023) 265203.

\bibitem{junker2023} G. Junker, S.H. Dong, P. Sedaghatnia, W.S. Chung, and H. Hassanabadi, {\it Ann. Phys.} \textbf{454} (2023) 169336.

\bibitem{junker2024} G. Junker, {\it J. Phys. A: Math. Theor.} \textbf{57} (2024) 075201.

\bibitem{schulze2024a} A. Schulze-Halberg, {\it Phys. Scr.} \textbf{99} (2024) 075212.

\bibitem{schulze2024b} A. Schulze-Halberg, and P. Roy, {\it J. Phys. A: Math. Theor.} \textbf{57} (2024) 225204.

\bibitem{schulze2024c} A. Schulze-Halberg, {\it Mod. Phys. Lett. A} \textbf{39} (2024) 2450178.

\bibitem{rabsaghari2024} A. Arabsaghari, H. Hassanabadi, and W.S. Chung, {\it Mod. Phys. Lett. A} \textbf{39} (2024) 2450117.

\bibitem{hassanabadi2024} S. Hassanabadi, J. K\v{r}\'i\v{z}, B.C. L\"utf\"uo\u{g}lu, W.S. Chung, P. Sedaghatnia, and H. Hassanabadi, {\it Int. J. Theor. Phys.} \textbf{63} (2024) 323.

\bibitem{benzair2024} H. Benzair, T. Boudjedaa, and M. Merad, {\it Phys. Scr.} \textbf{99} (2024) 055261.

\bibitem{raber2025} D.E.M. Raber, H. Benzair, T. Boudjedaa, and M. Merad, {\it Phys. Scr.} \textbf{100} (2025) 015277.

\bibitem{benarous2025} M. Benarous, A. Hocine, B.C. L\"utf\"uo\u{g}lu, and B. Hamil, {\it J. Stat. Mech.} \textbf{5} (2025) 053102.

\bibitem{hamil2025a} B. Hamil, B.C. L\"utf\"uo\u{g}lu, and M. Merad, {\it  Phys. Scr.} \textbf{100} (2025) 035301.

\bibitem{bouguerne2024a} H. Bouguerne, B. Hamil, B.C. L\"utf\"uo\u{g}lu, and M. Merad, {\it Indian J. Phys.} \textbf{98} (2024) 4093.

\bibitem{bouguerne2024b} H. Bouguerne,  B. Hamil, B.C. L\"utf\"uo\u{g}lu, and  M. Merad, {\it Nucl. Phys. B} \textbf{1007} (2024) 116684.

\bibitem{benchikha2024} A. Benchikha, B. Hamil, B.C. L\"utf\"uo\u{g}lu, and  B. Khantoul, {\it Int. J. Theor. Phys.} \textbf{63} (2024) 248.

\bibitem{lut2025} B.C. L\"utf\"uo\u{g}lu, A. Benchikha, B. Hamil, and B. Khantoul, {\it Mod. Phys. Lett. A} \textbf{40} (2025) 2550009.

\bibitem{hocine2025} A. Hocine, F. Merabtine, B. Hamil, B.C. L\"utf\"uo\u{g}lu, and  M. Benarous, {\it  Indian J. Phys.} \textbf{99} (2025) 775.

\bibitem{rou2023} N. Rouabhia, M. Merad, and B. Hamil, {\it  EPL} \textbf{143} (2023) 52003.

\bibitem{hamil2023} B. Hamil, B.C. L\"utf\"uo\u{g}lu, {\it  Physica A} \textbf{ 623} (2023) 128841.

\bibitem{hamil2025b}B. Hamil, B.C. L\"utf\"uo\u{g}lu, and M. Merad, {\it Mod. Phys. Lett. A} \textbf{40} (2025) 2550094.

\bibitem{salazar2026}M. Salazar-Ram\'irez, J.A. Mart\'inez-Nu\~no and M.R. Cordero-L\'opez, {\it Few-Body Syst.} \textbf{67} (2026) 3.






\bibitem{estes1968}L.E. Estes, T.H. Keil and L.M. Narducci, {\it Phys. Rev.} \textbf{175} (1968) 286.

\bibitem{caves1981}C.M. Caves, {\it Phys. Rev. D} \textbf{23} (1981) 1693.

\bibitem{gerry1987}C.C. Gerry, {\it Phys. Rev. A} \textbf{35} (1987) 2146.

\bibitem{chaturvedi1987}S. Chaturvedi, M.S. Sriram and V. Srinivasan, {\it J. Phys. A: Math. Gen.} \textbf{20} (1987) L1071.

\bibitem{gerry1989}C.C. Gerry, {\it Phys. Rev. A} \textbf{39} (1989) 3204.

\bibitem{vourdas1990}A. Vourdas, {\it Phys. Rev. A} \textbf{41} (1990) 1653.




\bibitem{villegas2018}B.M. Villegas-Mart\'inez, H.M. Moya-Cessa, and F. Soto-Eguibar, {\it Mod. Phys. Lett. B} \textbf{32} (2018) 1850247.

\bibitem{makarov2018} D.N. Makarov, {\it Phys. Rev. E} \textbf{97} (2018) 042203.

\bibitem{algarni2022} M. Algarni, K. Berrada, S. Abdel-Khalek, et al., {\it Mathematics} \textbf{10} (2022) 3051.

\bibitem{jones2023} C. Jones, J. Xavier, S.V. Kashanian, et al., {\it Opt. Express} \textbf{31} (2023) 10794.

\bibitem{nos2014} D. Ojeda-Guill\'en, R.D. Mota, and V.D. Granados, {\it J. Math. Phys.} {\bf55}, 042109 (2014).

\bibitem{choreno2018}E. Chore\~no, D. Ojeda-Guill\'en and V.D. Granados, {\it Eur. Phys. J. D} \textbf{72} (2018) 142.

\bibitem{vega2024}J.C. Vega, E. Chore\~no, D. Ojeda-Guill\'en and R.D. Mota, {\it J. Opt. Soc. Am. B} \textbf{41} (2024) 1084.

\bibitem{vega2025}J.C. Vega, D. Ojeda-Guill\'en and R.D. Mota, {\it J. Opt. Soc. Am. B} \textbf{42} (2025) 683.



\bibitem{moroz2016} A. Moroz, {\it EPL} \textbf{113} (2016) 50004.

\bibitem{hamil2022b} B. Hamil and B.C. L\"utf\"uo\u{g}lu, {\it Eur. Phys. J. Plus} \textbf{137} (2022) 812.

\bibitem{chung2021} W.S. Chung and H. Hassanabadi, {\it Few-Body Syst.} \textbf{62} (2021) 24.



\bibitem{nos2019b} E. Chore\~no, and D. Ojeda-Guill\'en, {\it Eur. Phys. J. Plus} {\bf134} 606 (2019).

\bibitem{nos2021b} E. Chore\~no, R. Valencia, and D. Ojeda-Guill\'en, {\it J. Math. Phys.} {\bf62} 071701 (2021).

\bibitem{nieto1997} M.M. Nieto, {\it Phys. Lett. A} {\bf229} 135 (1997).

\bibitem{mandel1979} L. Mandel, {\it Opt. Lett.} {\bf4} 205 (1979).

\bibitem{mandel1995} L. Mandel, and E. Wolf, {\it Optical Coherence and Quantum Optics}, Cambridge University Press, Cambridge, 1995.

\bibitem{glauber2007} R.J. Glauber, {\it Quantum Theory of Optical Coherence: Selected Papers and Lectures}, John Wiley \& Sons, Weinheim, 2007.

\bibitem{kim1989} M.S. Kim, F.A.M. de Oliveira, and P.L. Knight, {\it Phys. Rev. A} {\bf40} (1989) 2494.

\bibitem{walls1994} D.F. Walls, and G.J. Milburn, {\it Quantum Optics}, Springer, Berlin, 1994.


\end{thebibliography}
\end{document}